\documentclass[twocolumn,pre,preprintnumbers, showpacs, nofootinbib, superscriptaddress, amsmath,amssymb]{revtex4-1}
\usepackage{amstext,mathrsfs,amsthm,bm}
\usepackage{mathtools} % for symbols like := and =:
\usepackage{stmaryrd} % for variable-sized delimiters

\begin{document}

\newcommand{\smallfrac}[2]{\mbox{$\frac{#1}{#2}$}}

\title{Efficient computations of quantum canonical Gibbs state in phase space}

\author{Denys I. Bondar}
\email{dbondar@princeton.edu}
\affiliation{Princeton University, Princeton, New Jersey 08544, USA}

\author{Andre G. Campos}
%\email{agontijo@princeton.edu}
\affiliation{Princeton University, Princeton, New Jersey 08544, USA} 

\author{Renan Cabrera}
%\email{rcabrera@princeton.edu}
\affiliation{Princeton University, Princeton, New Jersey 08544, USA} 

\author{Herschel A. Rabitz}
%\email{hrabitz@princeton.edu}
\affiliation{Princeton University, Princeton, New Jersey 08544, USA} 

\date{\today}

\begin{abstract}
The Gibbs canonical state, as a maximum entropy density matrix, represents a quantum system in equilibrium with a thermostat. This state plays an essential role in thermodynamics and serves as the initial condition for nonequilibrium dynamical simulations. We solve a long standing problem for computing the Gibbs state Wigner function with nearly machine accuracy by solving the Bloch equation directly in the phase space. Furthermore, the algorithms are provided yielding high quality Wigner distributions for pure stationary states as well as for Thomas-Fermi and Bose-Einstein distributions. The developed numerical methods furnish a long-sought efficient computation framework for nonequilibrium quantum simulations directly in the Wigner representation.
\end{abstract}


\maketitle

\section{Introduction}

The state of a quantum system in thermodynamic equilibrium with a heat reservoir 
at temperature $T$ is given by the (un-normalized) Gibbs canonical density matrix
\begin{align}\label{GibsRhoDeffEq}
	\hat{\rho} = e^{-\beta\hat{H}}, \qquad \beta = 1/(kT),
\end{align}
where $\hat{H}$ is the quantum Hamiltonian. This state plays a fundamental role in  statistical mechanics. 
In particular, a system's equilibrium thermodynamical properties 
can be directly calculated from the corresponding
partition function
$ %\begin{align}
	Z = {\rm Tr}\, \left[ \hat{\rho} \right].
$ %\end{align}
Furthermore, studies of nonequilibrium dynamics driven by an external perturbation require 
the knowledge of the state in Eq. (\ref{GibsRhoDeffEq}), usually serving as both initial and final condition.

Recognizing the mathematical intractability for obtaining a generic quantum Gibbs state, 
Wigner attempted to derive a quantum correction to the corresponding classical ensemble by introducing 
the quasi probability distribution now bearing his name \cite{Wigner1932}. 
This discovery subsequently led to the development of  the phase space representation 
of quantum mechanics  \cite{Hillery1984121, zachos2005quantum, bolivar2004quantum, polkovnikov2010phase, Curtright2013, caldeira2014introduction}, where an observable $O=O(x,p)$ is a real-valued function of the 
coordinate $x$ and momentum $p$ [systems with one spatial dimension are considered; the scalability is discussed prior to Eq. (\ref{WignerOBMPSEq})], 
while the  system's state is represented by the Wigner function
\begin{align}\label{WignerFuncDeffEq}
	W_{xp} = 	\frac{1}{2\pi} \int  
		\langle  x - \smallfrac{\hbar}{2}\theta | \hat{\rho}  | x + \smallfrac{\hbar}{2}\theta  \rangle 
		e^{i p \theta}  d \theta,
\end{align}
where $\langle  x | \hat{\rho}  | x'  \rangle$ denotes a density matrix in the coordinate representation.

The Wigner function is a standard tool for studying the quantum-to-classical interface \cite{zurek2001sub,bolivar2004quantum, dragoman2004quantum, zachos2005quantum,Haroche2006, PhysRevD.58.025002, zachos2001, bolivar2004quantum, polkovnikov2010phase},  chaotic systems \cite{Heller84}, emergent classical dynamics \cite{Bhattacharya00, PhysRevLett.88.040402, Bhattacharya03, RevModPhys.75.715, Everitt09, jacobs2014quantum}, and open systems evolution \cite{Hillery1984121, Kapral2006, petruccione2002theory, Bolivar2012, caldeira2014introduction}. Moreover, the Wigner distribution has a broad range of applications in optics and signal processing \cite{Cohen1989, Dragoman2005, Schleich2001}, and  quantum computing \cite{Miquel2002a, Galvao2005, Cormick2006, Ferrie2009a, Veitch2012a, Mari2012, Veitch2013}.  Techniques for the experimental measurement of the Wigner function are also developed \cite{Kurtsiefer1997, Haroche2006, Ourjoumtsev2007, Deleglise2008, Manko2014}.

The knowledge of the Gibbs state Wigner function is essential for some models of nonequilibrium dynamics 
and transport phenomena (see, e.g., reviews \cite{coffey2007, Clearly2010, cleary_2011}). Despite numerous 
attempts, no \emph{ab initio} and universally valid method to obtain the Gibbs state in the phase space  exists. 
Based on recent analytical and algorithmic advances in the phase space representation 
of quantum mechanics \cite{PhysRevA.88.052108, Cabrera2015}, we finally deliver a numerically efficient 
and accurate method for calculating the Gibbs canonical state within the Wigner formalism. Additionally, 
a robust method to calculate ground and excited state Wigner functions are also designed. Thomas-Fermi and Bose-Einstein distributions in the Wigner representation are also computed. Since all the 
simulations presented below require the computational power of an average laptop, these algorithmic advancements 
enable quantum phase space simulations previously thought to be prohibitively expensive. 

The rest of the paper is organized as follows: The numerical method to calculate the Wigner function for the Gibbs canonical state is presented in Sec. \ref{Sec:NumMethodExplained}. Extensions of the algorithm to obtain the Wigner functions for pure stationary states as well as Thomas-Fermi and Bose-Einstein distributions are developed in Secs. \ref{Sec:GetPureStatesW}  and \ref{Sec:TFBE}, respectively. Python implementations of all the algorithms are supplied. Finally, the conclusions are drawn in the last section.

\section{Gibbs state as a Bloch equation solution}\label{Sec:NumMethodExplained}

Given the definition (\ref{WignerFuncDeffEq}), the problem of finding the Gibbs state Wigner function might appear 
to be trivial: substitute Eq. (\ref{GibsRhoDeffEq}) into Eq. (\ref{WignerFuncDeffEq}) and perform the numerical 
integration. However, this route is not only computationally demanding, but also yields poor results. In particular, 
the obtained function will not be a stationary state with respect to the dynamics generated by the Moyal equation 
of motion \cite{moyal1949quantum, zachos2005quantum, curtright2012quantum, Cabrera2015}
\begin{align}\label{MoyalEq}
	 i \hbar \, \partial_t W_{xp} = H_{xp} \star W_{xp} - W_{xp} \star H_{xp}, 
\end{align}
where $H_{xp}$ and $\hat{H}$ are connected via the Wigner transform (\ref{WignerFuncDeffEq}) and $\star$ denotes 
the Moyal product \cite{PhysRevD.58.025002, zachos2001, Curtright2013},
\begin{align}
	H_{xp} \star W_{xp} \equiv H_{xp} \exp \! \left( 
	  \smallfrac{i\hbar}{2} \, \overleftarrow{\partial_x} \, 
	  \overrightarrow{\partial_p} - 
	   \smallfrac{i\hbar}{2} \, 
	   \overleftarrow{\partial_p} \, \overrightarrow{\partial_x}
	  \right)  W_{xp},
\end{align}
which is a result of mapping the noncommutative operator product in the Hilbert space into the phase space. Note that
we follow the conventions of Ref. \cite{Cabrera2015} throughout. The Moyal equation (\ref{MoyalEq}) is obtained 
by Wigner transforming (\ref{WignerFuncDeffEq}) the von Neumann equation for the density matrix,
\begin{align}\label{vonNeumannEq}
	i \hbar \partial_t \hat{\rho} = [ \hat{H}, \hat{\rho} ].
\end{align}
Such a simple approach fails because the interpolation is required 
in Eq. (\ref{WignerFuncDeffEq}) to obtain values of the density matrix at half steps, as indicated 
by the $\hbar\theta/2$ shifts. Therefore, a different route must be taken to completely avoid the density matrix.

Following Refs. \cite{Hillery1984121, Clearly2010}, we note that the unnormalized  Gibbs 
state (\ref{GibsRhoDeffEq}) obeys the Bloch equation \cite{Bloch1932}
\begin{align}
	\partial_{\beta} \hat{\rho} = - (\hat{H}\hat{\rho} + \hat{\rho}\hat{H})/2,
	\qquad \hat{\rho}(\beta=0) = \hat{1}.
\end{align}
The latter could be written in the phase space as
\begin{align}\label{BlochPhaseSpaceEq}
 	\partial_{\beta} W_{xp} = -(H_{xp} \star W_{xp} + W_{xp} \star H_{xp})/2. 
 \end{align}
The Bloch equation in the Wigner representation is mathematically similar to the Moyal 
equation (\ref{MoyalEq}). Thus, a recently developed numerical 
propagator \cite{Cabrera2015} (as well as other methods \cite{thomann2016stability, machnes2016quantum,koda2015initial, koda2016mixed}) can be readily adapted to obtain the Gibbs state. 

Assume that the Hamiltonian is of the form
\begin{align}
	H_{xp} = K(p) + V(x).
\end{align}
To construct the numerical method, we first lift Eq. (\ref{BlochPhaseSpaceEq}) into the Hilbert 
phase space, as prescribed by Refs. \cite{PhysRevA.88.052108, Cabrera2015}, 
\begin{align}
     	\frac{d}{ d\beta}    |   \rho   \rangle &=  
	  -\frac{1}{2} \left[ H\left( \hat{x} - \smallfrac{\hbar}{2} \hat{\theta}, \hat{p} + \smallfrac{\hbar}{2} \hat{\lambda} \right)  \right. \notag\\
	 & \left. \qquad + H\left( \hat{x} + \smallfrac{\hbar}{2} \hat{\theta}, \hat{p} - \smallfrac{\hbar}{2} \hat{\lambda}\right) 
	   \right]    |   \rho  \rangle  \notag\\
	  &= -\frac{1}{2} \left[ \hat{K}^{+} + \hat{K}^{-} +  \hat{V}^{+} + \hat{V}^{-} \right] |   \rho   \rangle, \label{HilbertPhaseSpaceSchEq}\\
	  \hat{V}^{\pm} &= V\left(\hat{x} \pm \smallfrac{\hbar}{2} \hat{\theta} \right), \qquad
	  \hat{K}^{\pm} = K\left( \hat{p} \pm \smallfrac{\hbar}{2} \hat{\lambda}\right), \\
	  W_{xp} &= \smallfrac{1}{\sqrt{2 \pi \hbar}}  \langle x p  |  \rho \rangle,
\end{align}
where the four-operator algebra of self-adjoint operators $\hat{x},\hat{p}, \hat{\theta}, \hat{\lambda}$ satisfies the 
following commutator relations  \cite{bondar2012operational, PhysRevA.88.052108, Cabrera2015}:  
\begin{align} \label{commutation-rels}
 {[} \hat{x} , \hat{p} {]} = 0,  \quad
 {[} \hat{x} , \hat{\lambda} {]} = i, \quad
 {[} \hat{p} , \hat{\theta} {]} = i, \quad
 {[} \hat{\lambda} , \hat{\theta} {]} = 0,
\end{align}
and $|  x p \rangle$ denotes the common eigenvector of operators $\hat{x}$ and $\hat{p}$,
\begin{align}
	\hat{x} |  x p \rangle = x | x p \rangle, \qquad
	\hat{p} |  x p \rangle = p |  x p \rangle.
\end{align}  
The power of the Hilbert phase space formalism lies in the fact that the Bloch 
equation (\ref{BlochPhaseSpaceEq}) is transformed into Eq. (\ref{HilbertPhaseSpaceSchEq}) resembling 
an imaginary-time Schr\"odinger equation in two spatial dimensions, which could be efficiently 
solved via the spectral split operator method \cite{feit1982solution}. The formal 
solution of Eq. (\ref{HilbertPhaseSpaceSchEq}) reads
\begin{align}
	|   \rho(\beta)   \rangle
 = e^{-\frac{\beta}{2} \left( \hat{K}^{+} + \hat{K}^{-} +  \hat{V}^{+} + \hat{V}^{-} \right)} |   \rho(\beta=0)   \rangle.
\end{align}
Using the Trotter product~\cite{Trotter1959}, the iterative first-order scheme is obtained
\begin{align}
	|   \rho(\beta + d\beta )   \rangle =&
	e^{-\frac{d\beta}{2}\left(\hat{K}^{+} + \hat{K}^{-} \right)} \notag\\
	& \times e^{-\frac{d\beta}{2}\left( \hat{V}^{+} + \hat{V}^{-} \right)} |   \rho(\beta)   \rangle
	+ O\left( d\beta^2 \right).
\end{align}
Returning to the Wigner phase space representation, we finally arrive at the desired numerical scheme 
\begin{align}\label{BlochWignerNumerMethEq}
 W_{xp}(\beta + d\beta) &= \mathcal{F}^{\lambda \to x} 
 	e^{-\frac{d\beta}{2}\left(K^{+} + K^{-} \right)}
	\mathcal{F}^{x \to \lambda} \notag\\
	&\times \mathcal{F}_{\theta \to p}
  	 e^{-\frac{d\beta}{2}\left( V^{+} + V^{-} \right)} \mathcal{F}_{p \to \theta}  W_{xp}(\beta) , 
\end{align}
where $\mathcal{F}_{p \to \theta}$ and $\mathcal{F}^{x \to \lambda}$ are direct Fourier transforms with respect to the variables $p$ and $x$, respectively,
\begin{align}
 	\mathcal{F}_{p \to \theta}[W_{xp}] = \int W_{xp} e^{-ip\theta} dp, \\
	\mathcal{F}^{x \to \lambda}[W_{xp}] = \int W_{xp} e^{-ix\lambda} dx,
\end{align}
$\mathcal{F}_{\theta \to p}$ and $\mathcal{F}^{\lambda \to x}$ are the corresponding inverse transformations, and $V^\pm =  V\left( x \pm \smallfrac{\hbar}{2} \theta  \right)$, $K^{\pm} = K\left( p \pm \smallfrac{\hbar}{2} \lambda\right)$ have now become scalar functions.
Utilizing the fast Fourier transforms \cite{frigo2005design}, the complexity 
of the algorithm (\ref{BlochWignerNumerMethEq}) is $O(N\log N)$, where $N$ is the total length of an array storing the Wigner function. Moreover, the Wigner function at every iteration corresponds to a 
Gibbs state of a certain temperature, and Eq. (\ref{BlochWignerNumerMethEq}) physically models cooling.

In the current work, we consider one-body systems. The \emph{ab initio} algorithm (\ref{BlochWignerNumerMethEq}) can be straightforwardly extended to the $D$-body case albeit at the price of the exponential scaling $O\left(DN^D \log N\right)$. In the subsequent work, we will present a polynomial algorithm by adapting the matrix product state formalism \cite{wall_out_equilibrium_2012, orus_practical_2014} to  phase space dynamics. For example, deployment of the following matrix product state ansatz for a $D$-body Wigner function, $W^{(D)} = W\left(x_1, p_1; x_2, p_2; \ldots ; x_D, p_D\right)$,
\begin{align}\label{WignerOBMPSEq}
	W^{(D)} = \prod_{k=1}^{D-1} W_k (x_k, p_k; x_{k+1}, p_{k+1})
\end{align}
should lead to the desired polynomial scaling.

\begin{figure}
	\includegraphics[width=1.\hsize]{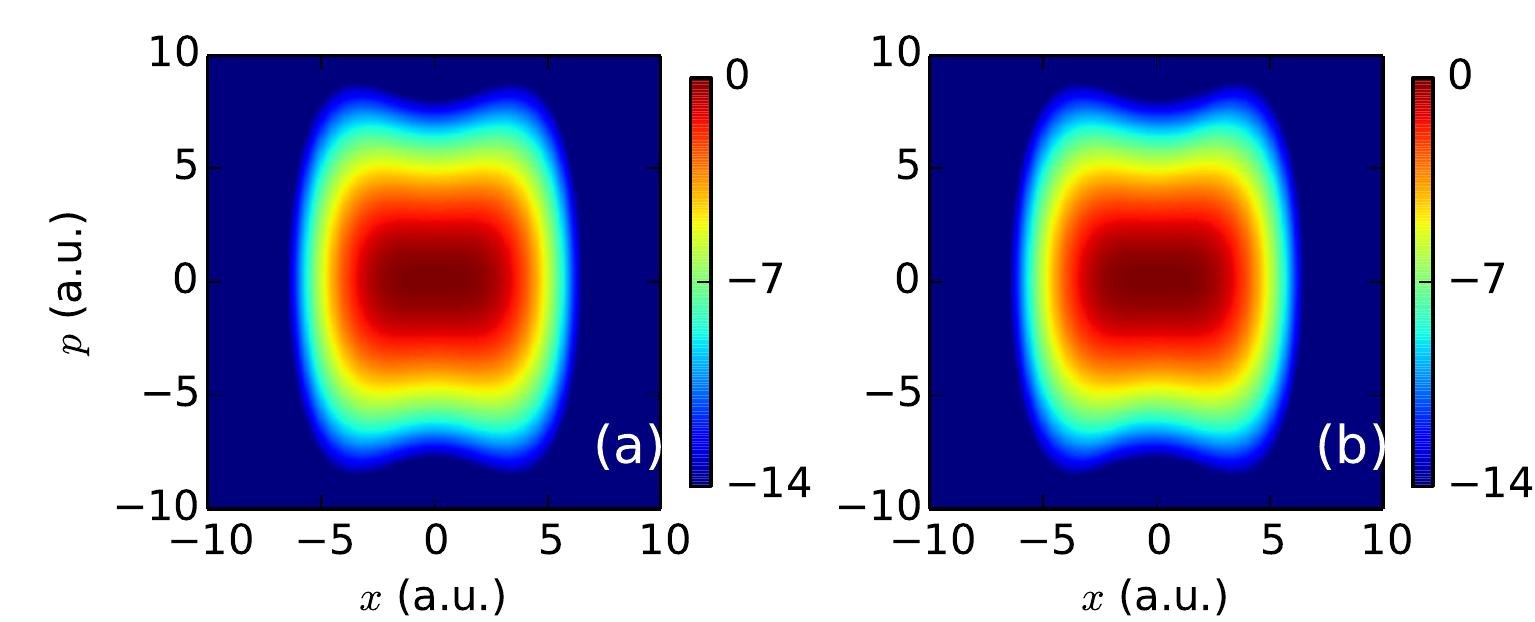}
 	\caption{(Color online) Log plot of the Gibbs canonical state Wigner function ($\beta = 1$ a.u.) for the Mexican hat system with the Hamiltonian $H_{xp} = p^2/2 -0.05x^2 + 0.03x^4$ (a.u.) Since the Gibbs state is characterized by a positive Wigner function, we use the logarithmic scale to show that the Gibbs distribution obtained by Eq. (\ref{BlochWignerNumerMethEq}) [Fig. (a)] remains invariant under the time evolution of the Moyal equation (\ref{MoyalEq}) up to the values of $10^{-14}$. See Ref. \cite{CodeFig1} regarding the python code used to generate this figure.}\label{fig:MexicanHatGibbsState}
\end{figure}

In Fig. \ref{fig:MexicanHatGibbsState}, we employ Eq. (\ref{BlochWignerNumerMethEq}) to compute the Gibbs state Wigner 
function for a Mexican hat potential. Atomic units (a.u.), where $\hbar=m=1$, are used throughout. To verify the consistency 
of the obtained solution, we subsequently propagate it by the Moyal equation (\ref{MoyalEq}) using the method 
in Ref. \cite{Cabrera2015}. Comparing the initial [Fig. \ref{fig:MexicanHatGibbsState}(a)] 
and final [Fig. \ref{fig:MexicanHatGibbsState}(b)] states, one observes that the Gibbs state remains stationary 
up to $O\left( 10^{-14} \right)$.

\begin{figure}
	\includegraphics[width=1.\hsize]{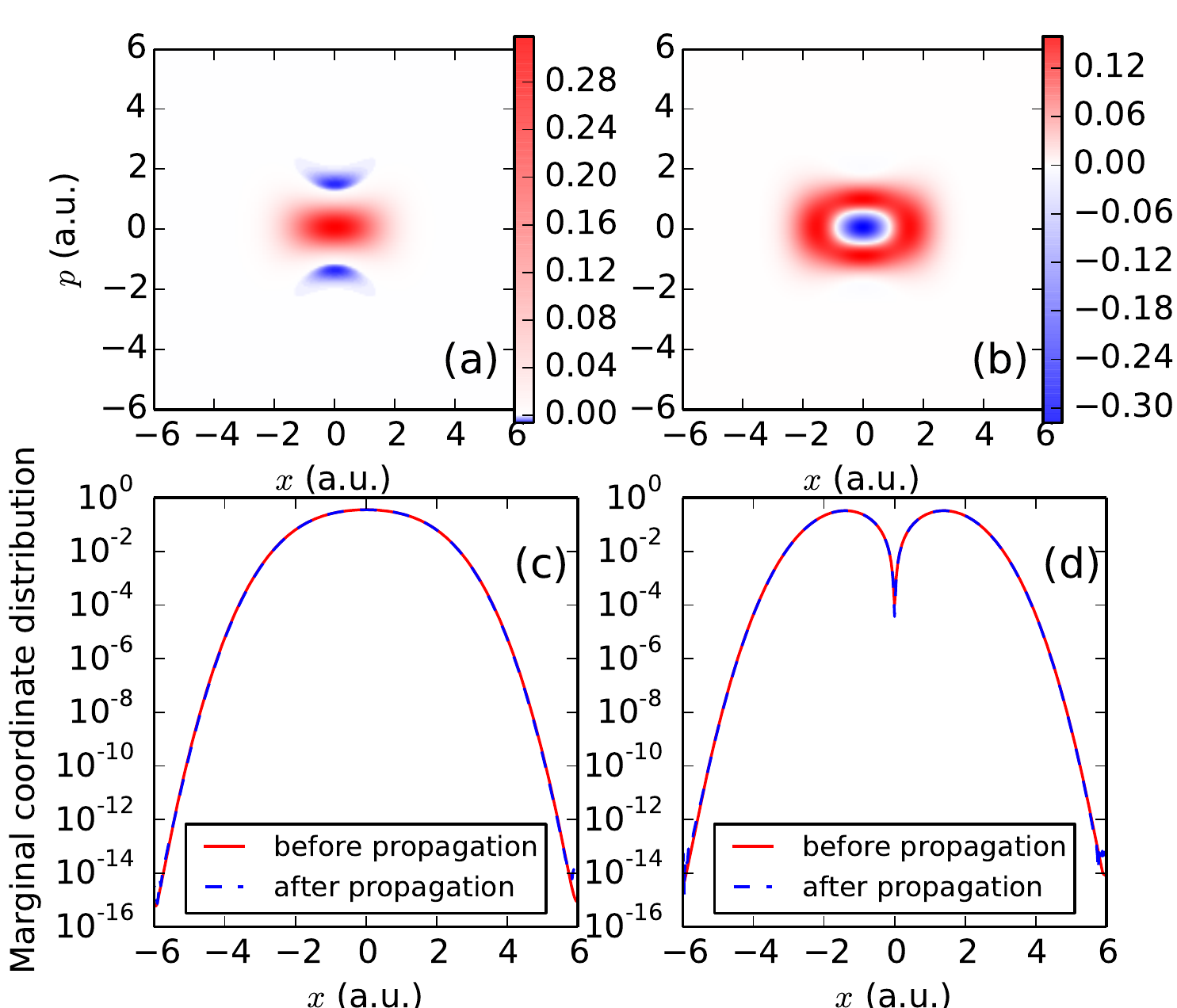}
 	\caption{(Color online) Wigner functions of the ground (a) and first excited states (b) for the Mexican hat system with the Hamiltonian $H_{xp} = p^2/2 -0.05x^2 + 0.03x^4$ (a.u.). In Figs. (c) and (d), the red solid line depict the marginal coordinate distribution of the ground and first excited state, respectively; whereas, the dashed blue line depict the coordinate distribution obtained after the propagation via the Moyal equation (\ref{MoyalEq}). Note that both lines overlap, indicating that the pure states in Figs. (a) and (b) are calculated with high accuracy. See Ref. \cite{CodeFig2} regarding a python code used to generate this figure.}\label{fig:MexicanHatGroundExitedFig}
\end{figure}

\section{Wigner functions of pure stationary states}\label{Sec:GetPureStatesW}

The numerical scheme (\ref{BlochWignerNumerMethEq}) recovers the ground state as $\beta\to\infty$. To speed up the convergence 
to the zero-temperature ground state, the following adaptive step algorithm can be employed.
Initially pick a large value of the inverse temperature step $d\beta \sim 1$ (a.u.). Using a constant Wigner 
function [i.e., $W_{xp}(\beta=0)=1$] as an initial guess, obtain $W_{xp}(\beta + d\beta)$ within Eq. (\ref{BlochWignerNumerMethEq}). 
Accept the updated Wigner function, if it lowers the energy [i.e.,
$
	\int W_{xp}(\beta) H_{xp} dxdp > \int W_{xp}(\beta + d\beta) H_{xp} dxdp
$]
 and $W_{xp}(\beta + d\beta)$ represents a physically valid state. If either condition is violated, reject the state $W_{xp}(\beta + d\beta)$, 
half the increment $d\beta$, and repeat the procedure. 
 
Note that there is no computationally feasible criterion to verify that a Wigner function underlines 
a positive density matrix (see, e.g., Ref. \cite{Ganguli1998}). Thus, we suggest to employ the following 
heuristic: verify that the purity,
$
	\mathcal{P} = 2\pi\hbar \int W_{xp}^2 dxdp
$ 
cannot exceed unity (note that $\mathcal{P} =1$ is for pure states only) and the Heisenberg uncertainty principle is obeyed. 
See Ref. \cite{CodeFig2} regarding a pythonic implementation of the full algorithm.

Once the ground state is found, any exited state can be constructed in a similar fashion. For example, 
an amended algorithm can be used to calculate the first exited state. After Eq. (\ref{BlochWignerNumerMethEq}), the 
ground state Wigner function $W_{xp}^{(g)} \coloneqq W_{xp}(\beta=\infty)$ should be projected  
  out of $W_{xp}(\beta + d\beta)$. More specifically, the state $W_{xp}(\beta + d\beta)$ must be updated as
\begin{align}
	W_{xp}(\beta + d\beta) &\coloneqq \frac{W_{xp}^{(1)}}{ \int W_{xp}^{(1)} dxdp},\\
	W_{xp}^{(1)} &= W_{xp}(\beta + d\beta) 
	- c W_{xp}^{(g)} , \\
	c &= 2\pi\hbar \int W_{xp}(\beta + d\beta) W_{xp}^{(g)} dx dp.
\end{align}
The physical meaning of $c$ is a fraction of the total population occupying the ground state. This algorithm, implemented in Ref. \cite{CodeFig2}, is more efficient and easier to maintain than the one in Refs. \cite{hug_1998a, hug_1998b}.

Figures \ref{fig:MexicanHatGroundExitedFig}(a) and  \ref{fig:MexicanHatGroundExitedFig}(b) show the Wigner functions 
of ground and first exited states, respectively, for the Mexican hat potential. Using merely $512 \times 512$ grids to store 
Wigner functions, the purities of the computed states are found to be $1 - O\left( 10^{-14} \right)$ and $1- O\left( 10^{-7} \right)$, 
respectively. This demonstrates numerical effectiveness of the developed algorithm. Furthermore, to verify that 
the states are stationary, we propagated them via the Moyal equation (\ref{MoyalEq}) \cite{Cabrera2015}. The 
comparison of the coordinate marginal distributions, defined as $\int W_{xp}dp$, before and after 
Moyal propagation [Figs. \ref{fig:MexicanHatGroundExitedFig}(c) and \ref{fig:MexicanHatGroundExitedFig}(d)] confirm 
that the ground [Fig. \ref{fig:MexicanHatGroundExitedFig}(a)] and exited [Fig. \ref{fig:MexicanHatGroundExitedFig}(b)] 
states are stationary within the accuracy of $O\left( 10^{-14} \right)$. 

It is noteworthy that the ground state Wigner function [Fig. \ref{fig:MexicanHatGroundExitedFig}(a)] exhibits small negative values. 
This is in compliance with Hudson's theorem \cite{Hudson1974} stating that a pure state Wigner function is positive 
if and only if the underlying wave function is a Gaussian (whereas, the ground state of the Mexican hat potential 
is evidently non-Gaussian). The excited state Wigner function [Fig. \ref{fig:MexicanHatGroundExitedFig}(b)] contains 
a central oval region with pronounced negative values encircled by a zero-valued oval followed by a positive-value region. 
This is a hallmark structure of the Wigner distributions for first excited states. 
The zero-valued oval and negative center emerge from the node at $x=0$ in the first excited state 
wave function, which can also be seen in Fig. \ref{fig:MexicanHatGroundExitedFig}(d) visualizing the absolute 
value square of the wave function. 
Note that Wigner function's negativity is associated with the exponential speedup in quantum computation \cite{Ferrie2009a, Veitch2012a, Mari2012, Veitch2013}.

\section{Wigner functions for Thomas-Fermi and Bose-Einstein distributions}\label{Sec:TFBE}

\begin{figure}
	\includegraphics[width=1\hsize]{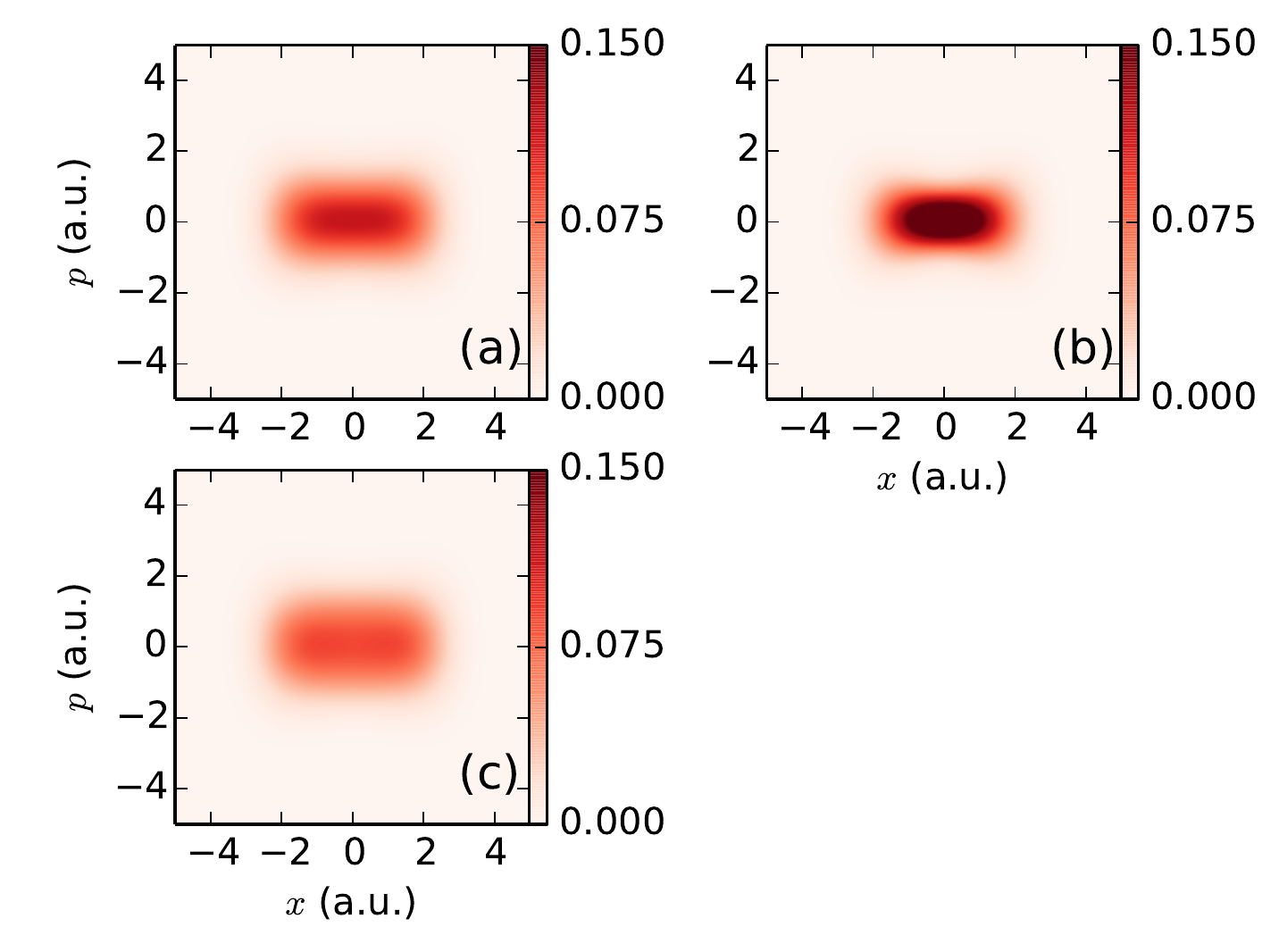}
 	\caption{(Color online) Wigner functions of the (a)  Gibbs, (b)  Bose-Einstein, and (c) Thomas-Fermi states for the Mexican hat system with the Hamiltonian $H_{xp} = p^2/2 -0.05x^2 + 0.03x^4$ (a.u.),  $\beta = 1.5$ (a.u.), and $\mu = 0$. Distributions (b) and (c) were obtained using expansion (\ref{TFBEWignerEq}). See Ref. \cite{CodeFig3} regarding a python code used to generate this figure.
	}\label{fig:MexicanHatBE_TF}
\end{figure}

The proposed method can be used to compute other steady states not directly describable by the Bloch equation. In particular,
to calculate the Thomas-Fermi ($s=+1$) or Bose-Einstein ($s=-1$) states, the following expansion can be utilized
\begin{align}\label{TFBEWignerEq}
	& \frac{1}{ e^{\beta(\hat{H} - \mu)} + s }
		= \frac{e^{-\beta(\hat{H} - \mu)}}{ 1 + s e^{-\beta(\hat{H} - \mu)} }  \notag\\
	& = \sum_{k=0}^{\infty} e^{(2k+1)\beta \mu} e^{-(2k+1)\beta \hat{H}}  
			 -s \sum_{k=1}^{\infty} e^{(2k)\beta\mu} e^{-(2k)\beta\hat{H}},
\end{align}
where $\mu$ denotes the chemical potential. Equation (\ref{TFBEWignerEq}) consists of the (unnormalized) Gibbs states at different temperatures $\beta, 2\beta, 3\beta, \ldots$. Thus the Wigner function for the Thomas-Fermi and Bose-Einstein states could be easily found via the numerical method (\ref{BlochWignerNumerMethEq}) by adding or subtracting the corresponding Gibbs distributions, which are sequentially obtained during $\beta \to \infty$ propagation. In Fig. \ref{fig:MexicanHatBE_TF}, the Gibbs state [Fig. \ref{fig:MexicanHatBE_TF}(a)] has been compared with the Bose-Einstein [Fig. \ref{fig:MexicanHatBE_TF}(b)] and Thomas-Fermi  [Fig. \ref{fig:MexicanHatBE_TF}(c)] distributions for $\beta = 1.5$ (a.u.) and vanishing chemical potential.

\section{Outlook}

The Gibbs canonical state, a maximum entropy stationary solution of the von Neumann equation (\ref{vonNeumannEq}), 
is a cornerstone of quantum statistical mechanics. The Wigner phase space representation of quantum dynamics 
currently undergoes a renewing  interest due to the promise to solve open problems in nonequilibrium thermodynamics. To simulate open system dynamics, a good quality initial condition, usually the Gibbs 
state Wigner function, needs to be supplied. We have developed the numerical algorithm yielding Gibbs states with 
nearly machine precision. Moreover, an extension of this algorithm allows computing Wigner functions of pure 
stationary states, corresponding to the eigensolutions of the Schr\"{o}dinger equation. Wigner functions for Thomas-Fermi and Bose-Einstein distributions are also calculated. Such states are essential 
for studying nonequilibrium dynamics in atomic and molecular systems. As a result, the developed algorithmic techniques
finally make the Wigner quasiprobability phase space representation of quantum dynamics 
a  computationally advantageous formulation compared to the density matrix approach. 

\emph{Acknowledgments}. The authors acknowledge financial support 
from (H.A.R.) NSF CHE 1058644, (R.C.) DOE DE-FG02-02-ER-15344 and (D.I.B.) ARO-MURI W911-NF-11-1-2068. A.G.C. was supported by the Fulbright foundation. D.I.B. was also supported by 2016 AFOSR Young Investigator Research Program. 
\bibliography{QuantumGibbsState}
\end{document}